\documentclass[12pt]{article}

\usepackage{amsmath,amssymb}
\usepackage{graphicx}
\usepackage{caption}
\usepackage{color}
\usepackage{dcolumn}
\usepackage{bm}
\usepackage{float}
\usepackage{hyperref} 
\usepackage{natbib}
\usepackage{verbatim}
\usepackage{bm}
\hypersetup{ colorlinks = true, citecolor = black, urlcolor = blue}

\definecolor{background-color}{gray}{0.98}

\newcommand{\ds}{\displaystyle}
\newcommand{\E}{\text{E}}
\newcommand{\Var}{\text{Var}}
\newcommand{\X}{\mathcal{X}}

\newcommand*\samethanks[1][\value{footnote}]{\footnotemark[#1]}

\title{Analyzing MCMC output}
\date{\today}

\author{Dootika Vats\thanks{Department of Mathematics and Statistics,
    Indian Institute of Technology Kanpur}, Nathan
  Robertson\thanks{Department of Statistics, University of California,
    Riverside}, James M. Flegal\samethanks, Galin L. Jones
  \thanks{School of Statistics, University of Minnesota, Twin-Cities.
  Partially supported by National Science Foundation award 1922512.}}

\begin{document}

\maketitle

\begin{center}
\subsubsection*{\small Article Type:}
Overview

\hfill \break
\thanks

\subsubsection*{Abstract}
\begin{flushleft}
  Markov chain Monte Carlo (MCMC) is a sampling-based method for
  estimating features of probability distributions. MCMC methods produce a serially correlated, yet representative, sample from
  the desired distribution.  As such it can be difficult to assess when
  the MCMC method is producing reliable results.  We present some
  fundamental methods for ensuring a reliable simulation
  experiment.  In particular, we present a workflow for output
  analysis in MCMC providing estimators, approximate sampling
  distributions, stopping rules, and visualization tools.
\end{flushleft}
\end{center}

\clearpage

\renewcommand{\baselinestretch}{1.5}
\normalsize

\clearpage

\section{Introduction} 
\label{sec:introduction}

Markov chain Monte Carlo (MCMC) algorithms are an essential tool for
estimating features of probability distributions encountered in diverse
applications \citep{broo:gelm:jone:meng:2011}.  The use of MCMC is commonly
identified with Bayesian settings, but it is also useful in other situations
\citep[see e.g.][]{caff:jank:jone:2005, geyer:1991, gjoka:crawl:2011}. 

Suppose, for our application, we have developed a probability
distribution $F$ with support $\mathcal{X} \subseteq \mathbb{R}^d$,
$d \ge 1$.  Our goal is to use fixed, unknown features of $F$ to make
inference about the population. For example, for
$h : \mathcal{X} \to \mathbb{R}$, we may be interested in the
expectation\footnote{If $f$ is a probability function, then this notation avoids
  having separate formulas for the continuous case where
  $\mu_{h} = \int_{\mathcal{X}} h(x) f(x) dx$ and the discrete case
  where $\mu_{h} = \sum_{x \in \mathcal{X}} h(x) f(x)$ but it also
  applies to more general settings.}
\[  
\mu_h = \E_{F}[h(X)] = \int_{\mathcal{X}} h(x) F(dx) \; .   
\]

The apparent simplicity of this hides an array of features, including
probabilities, means, moments, and marginal densities associated with
$F$.  Accordingly, we will typically want to use several expectations.
There are features of $F$ that are not expectations, such as
quantiles, but we will defer discussion of this to
Section~\ref{sec:estimation_process}.  We collect all the features
of $F$ we want in a $p$-dimensional vector, $\theta$.

We use the notation $\sim$ to mean ``distributed as,''
$\stackrel{{\boldsymbol\cdot}}{\sim}$ to mean ``approximately
distributed as,'' $\not \sim$ to mean ``not distributed as,'' and
$\approx$ to mean ``is approximately equal to.'' Often $F$ is
analytically intractable in the sense that directly calculating
$\theta$ is impossible and hence we may turn to estimating $\theta$
using Monte Carlo sampling methods.  We consider only MCMC in which a
realization of a (aperiodic, Harris recurrent -- for definitions see
\cite{meyn:twee:2009}) Markov chain $\{X_1, X_2, \ldots, X_n\}$ is
produce so that for a sufficiently large Monte Carlo sample size $n$,
we have $X_n \stackrel{{\boldsymbol\cdot}}{\sim} F$. We will not
discuss how to construct or implement MCMC methods
\citep[see][]{chib:gree:1995, robe:case:2013,
  robert:elvira:tawn:wu:2018}, but will instead assume that there is
already an efficient method for producing the MCMC sample, or output.
The output is then used to construct estimators of $\theta$ so that
for a sufficiently large Monte Carlo sample size, $n$, we have
\[ 
\hat{\theta}_{n} = \hat{\theta}(X_1, \ldots, X_n) \approx \theta .  
\]
In typical implementations of MCMC, the initial state, $X_1$,  is drawn from some other distribution than the target, that is, $X_1 \nsim F$, which
implies $X_n \nsim F$ for any finite Monte Carlo sample size $n$.
Moreover, there is an inherent serial correlation in the MCMC
sample. That is, the Markov chain draws are neither independent nor
identically distributed.  Thus, there are two common tasks in MCMC
output analysis (i) deciding when the simulation has produced a
representative sample from $F$, that is, when is $n$ large enough so
that $X_n \stackrel{{\boldsymbol\cdot}}{\sim} F$, and (ii) when is $n$
sufficiently large to conclude $\hat{\theta}_{n} \approx \theta$.
Notice that the required number of draws may be different for each
task.  Task (i) is difficult and while there are some rigorous
approaches \citep{rose:1995a}, these are typically challenging to
implement in practically relevant settings.  This has led most
practitioners to approach this question by relying on graphical
summaries and ad hoc convergence diagnostics; see
Section~\ref{sec:burn_in}.  Task (ii) is our main focus and is
fundamental to ensuring a reliable simulation experiment.  Classical
large-sample frequentist methods provide principled, practical
solutions for (ii).  However, since it is a Markov chain that is being
simulated, specialized techniques are required for their
implementation; this is covered in some detail in
Sections~\ref{sec:estimation_process}--\ref{sec:ext}.

In practice, addressing tasks (i) and (ii) can seem complicated to the
uninitiated.  Thus, we present a workflow for analyzing output from
MCMC addressing these challenges.  This is illustrated in the context of a
Bayesian example in Section~\ref{sec:example}. 

\section{Markov chain theory and starting values} 
\label{sec:burn_in}
The dynamics of a time-homogeneous Markov chain are dictated by its
Markov transition kernel
\[
P^n(x, A) := \Pr(X_{n+j} \in A \mid X_{j} =x)\,,  ~~~~ n,j\ge 1
\]
and $P^1 \equiv P$. If $\nu$ is the initial distribution (i.e.\
$X_1 \sim \nu$), then $\nu P^n$ is the marginal distribution of
$X_n$. Markov chains for MCMC are constructed in such a way that the
target distribution $F$, is its stationary distribution. That is,
$FP^n = F$, so that if $X_1 \sim F$, then each $X_n \sim F$.  Of
course, producing $X_1 \sim F$ often is not possible in settings where
MCMC is relevant. However, if $\| \cdot \|$ denotes total variation
distance, under standard conditions \citep[for an accessible
discussion see][]{jone:hobe:2001, robe:rose:2004},
\begin{equation}
\label{eq:tv}
\|\nu P^n - F \| \to 0 \text{ as $n \to \infty$}.
\end{equation}
This implies $X_n \stackrel{d}{\to} F$ as $n \to \infty$.  Moreover,
the total variation norm in \eqref{eq:tv} is non-increasing in $n$
\citep{meyn:twee:2009}.  Hence a representative, although correlated,
sample will be produced eventually, that is, for sufficiently large $n$.

Ideally, we would like to identify $n^\star < \infty$ such that if
$n \ge n^\star$, then $\|\nu P^n - F \| < \epsilon$ for some
$\epsilon >0$ so that we could conclude
$X_n \stackrel{{\boldsymbol \cdot}}{\sim} F$.  Indeed, there are
methods for doing so \citep{jone:hobe:2001, rose:1995a}, but they are
often so conservative as to be of little practical use or are
difficult to apply \citep{jone:hobe:2004}.  This has forced
practitioners to turn to other methods, the most common of which are
trace plots of the components of the simulation and so-called
convergence diagnostics \citep{cowl:carl:1996}.  Perhaps the most
commonly used convergence diagnostic was developed by
\cite{gelm:rubi:1992a}.  However, this diagnostic has been shown to
have severe limitations \citep{fleg:hara:jone:2008, veht:gel:2019};
see Section~\ref{sec:stopping_mcmc} for more. In fact, many
convergence diagnostics, including the Gelman-Rubin diagnostic, were
developed to address task (ii) in Section~\ref{sec:introduction}, but
are often incorrectly understood to answer task (i).  In general, all
convergence diagnostics should be used with care since their very use
can introduce bias \citep{cowl:robe:rose:1999}.  Strictly speaking,
convergence does not occur for any finite $n$ so all they can detect
is evidence of non-convergence.  Hence diagnostics can never guarantee
what we want because absence of evidence of non-convergence is not
evidence of convergence. For more discussion on convergence
diagnostics see \cite{roy:2019}.

While the convergence in \eqref{eq:tv} holds for any starting value,
some will be better than others since the rate of convergence can be
affected by the choice of starting value \citep{rose:1995a}. In
particular, starting in an area of low probability of $F$ can lead to
slow convergence of the Markov chain \citep{gilk:rich:spie:1996}. If,
however, $X_1\sim F$, then $\|FP^n - F\| = 0$ for all $n$, and the
Markov chain produces exact draws from $F$ (albeit still
correlated). Thus, starting values can have a substantial impact on
the quality of the samples. Choosing good starting values can save
considerable time in postprocessing and increase confidence in the
results. 

It is a truism that ``Any point you don't mind having in a sample is a
good starting point.'' \citep{geye:2011}. While choosing a good
starting value may be difficult, it may not be impossible. In fact, it
may be possible to start from stationarity via perfect simulation
\citep{huber:2016}, Bernoulli factories \citep{fleg:herb:2012}, or
simple accept-reject samplers. When the target distribution is
low-dimensional or made low-dimensional by a linchpin variable trick
\citep{arch:2016}, an accept-reject sampler that can produce one 
sample from the target may be available. Such a sampler is often too
computationally burdensome for a full Monte Carlo procedure, but may
provide a single draw at reasonable cost.

Starting values are also often obtained by finding a high probability
region via optimization.  When a closed-form expression is available, using the maximum
likelihood estimate can be computationally cheap.  Other optimization approaches are  certainly possible, but
finding a global optimum is in general difficult.  Even so, in many
situations, any value in a high probability region is a reasonable
starting value.  In Bayesian settings, practitioners may draw starting
values from the (proper) prior distributions of the parameters.
This can work particularly well when the prior distributions have been
chosen with care.  Finally, and particularly when implementing
component-wise MCMC methods \citep{john:jone:neat:2013} such as Gibbs
samplers, the parameter being updated first may not require a starting
value. So it is often a good idea to first update the parameter whose
starting value is least trustworthy.

\section{Estimation and sampling distributions} 
\label{sec:estimation_process} Recall that given a realization
$\{X_1, X_2, \dots, X_n\}$ of the Markov chain we construct an
estimator of $\theta$,
$\hat{\theta}_{n} = \hat{\theta}(X_1, \ldots, X_n)$ so that
$\hat{\theta}_{n} \to \theta$ almost surely as $n \to \infty$.
However, no matter how large the (finite) Monte Carlo sample size $n$,
there will be an unknown Monte Carlo error
$\hat{\theta}_{n} - \theta$.  The approximate sampling distribution of
the Monte Carlo error is often available through a version of the
central limit theorem (CLT), which holds under moment conditions on
the functionals and Markov chain mixing conditions \citep{jone:2004},
both of which require theoretical study to verify in a given
application.  \citet{jone:hobe:2001} give an accessible introduction to
this theory which has been applied in a number of practically
relevant settings \citep[see e.g.][]{ekva:jone:2019,
  hobe:jone:pres:rose:2002, john:jone:2015, khare:hobe:2013,
  lund:twee:1996, robe:twee:1996,
  roy:hobe:2007, roy:hobe:2010, tan:hobe:2012, vats:2017}

\subsection*{Means} 
\label{sub:means}
In most settings we will want to estimate several expectations.  Let
$h: \mathcal{X} \to \mathbb{R}^p$ be a function such that
$\E_Fh(X)= \mu_{h}$ is of interest. For example, to estimate the mean
vector of $F$, $h$ will be the identity function. Estimation is
straightforward using a
sample mean since the Markov chain strong law ensures
\[
\bar{\mu}_n:= \dfrac{1}{n} \ds \sum_{t = 1}^{n} h(X_t) \overset{a.s.}{\to} \mu_{h}\quad \text{as $n \to \infty$}\,.
\]
If a CLT holds, then there exists a $p \times p$ positive definite matrix,
$\Sigma$, such that as $n\to \infty$,
\[
\sqrt{n} \left(\bar{\mu}_n - \mu_{h} \right) \overset{d}{\to} N_p(0, \Sigma)\,.
\]
Here, $\Sigma$ encodes the covariance structure for $h$ in the target distribution and the serial lag-covariance due to the Markov chain. More specifically,
\begin{equation}
\label{eq:sigma_clt}
\Sigma = \ds \sum_{k=-\infty}^{\infty} \text{Cov}_F(h(X_1), h(X_{1+k}))\,.	
\end{equation}
The subscript $F$ in \eqref{eq:sigma_clt} means that the expectations
are calculated under the assumption that $X_1 \sim F$.  This does
\emph{not} mean that we need $X_1 \sim F$ for the CLT to hold.
Indeed, if the CLT holds for one initial distribution, then it holds
for every initial distribution \citep{meyn:twee:2009}.

Under an independent sampling scheme,
$\text{Cov}_F(h(X_1), h(X_{1+k})) = 0$ for all $k \ne 0$, but the
Markov chains encountered in MCMC applications exhibit  serial
dependence. Thus, utilizing the sampling distribution for
$\bar{\mu}_n$ to make large-sample inference requires specialized
methods for estimating $\Sigma$, which we discuss later.

\subsection*{Quantiles} 
\label{sub:quant}
In addition to expectations, marginal quantiles are often of interest. Let
$h:\X \to \mathbb{R}$, and for $X \sim F$, set $V = h(X)$. Let $F_V$ be the
distribution of $V$, and for $0 < q < 1$, the $q$-quantile of $F_V$ is defined
as:
\[
\phi_{q} = \inf\{x: F_V(x) \geq q  \}\,.
\]
A natural estimator of $\phi_{q}$ is the $\lceil nq \rceil$th order statistic. If $\{V_1,
V_2, \dots, V_n\}$ is the transformed process, and $\{V_{(1)}, V_{(2)}, \dots,
V_{(n)}\}$ are the order statistics, then an estimator of $\phi_{q}$ is
\[
\hat{\phi}_{q}:= V_{(j+1)} \text{ where } j < nq  \leq  j+1\,,
\]
and $\hat{\phi}_q \overset{a.s.}{\to} \phi_q$ as $n \to \infty$.  An
approximate sampling distribution for $\hat{\phi}_q$ is developed by
\cite{doss:fleg:jone:neat:2014}. First, for any $y$ define, \[
\sigma^2(y) =  \sum_{k=-\infty}^{\infty} \text{Cov}_{F_V}(I (V_1 \leq y), \, I (V_{1+k} \leq y))\,,
\]
and let $f_{V}$ be the density associated with $F_V$. Then, as $n \to \infty$,
\[
\sqrt{n} (\hat{\phi}_q - \phi_q) \overset{d}{\to}N(0, \sigma^2(\phi_q)/f_V(\phi_q)^2)\,.
\]
Here, we present a univariate sampling distribution for the quantiles, but
often multiple quantiles may be of interest. The joint sampling distributions
of multiple quantiles and of means and quantiles is available in
\cite{robe:visual:2019}.

\subsection*{Functions of expectations} 
\label{sub:other}

We are often interested in estimating functions of expectations, in
which case a plug-in type estimator is natural, and the approximate
sampling distribution may be obtained by an application of the delta
method. This includes estimating the covariance matrix of $F$,
$\Lambda = \Var_F(h(X_1))$, with the sample covariance matrix,
\[
\Lambda_n := \dfrac{1}{n}\ds \sum_{t = 1}^{n} (h(X_t) -
\bar{\mu}_n)(h(X_t) - \bar{\mu}_n)^T .
\]
A delta method argument similar to the independent and identically
distributed setting yields an element-wise sampling distribution for
$\Lambda_n$.

\section{Estimating Monte Carlo error} 
\label{sec:estimating_sigma}

The approximate sampling distributions of the previous section provide
the keys to assessing the reliability of the simulation effort, that
is, addressing task (ii) from
Section~\ref{sec:introduction}. Construction of confidence regions for
$\mu_h$ to address this problem has attracted substantial interest
\citep{atch:2016, fleg:gong:2015, jone:hara:caff:neat:2006,
  rosen:2017, vats:fleg:jones:2019}.  Suppose that $\Sigma_{n}$ is an
estimator of $\Sigma$ in the CLT \eqref{eq:sigma_clt}.  If
$T^{2}_{1-\alpha, p , q}$ denotes a $1-\alpha$ quantile from a
Hotelling's $T$-squared distribution with dimensionality $p$ and
degrees of freedom $q$, then it is straightforward to construct an
asymptotic $100(1-\alpha)\%$ confidence region
\[ 
C_{\alpha}(n) = \left\{ n(\bar{\mu}_{n} - \mu_{h})^T \Sigma_{n}^{-1}
  (\bar{\mu}_{n} - \mu_{h}) < T^{2}_{1-\alpha, p , q} \right\}  . 
\]
The size of $C_{\alpha}(n)$ will then describe the precision of
estimation; we will discuss how to use this information in the
sequel. Thus, the main obstacle is estimating the variance in the
asymptotic distribution. There are a variety of estimators available
that broadly fit into three classes (1) spectral variance estimators
\citep{andr:1991, dame:1991, fleg:jone:2010, vats:fleg:jones:2018},
(2) batch means estimators \citep{chen:seila:1987,
  liu:fleg:2018, jone:hara:caff:neat:2006, vats:fleg:jones:2019}, and
(3) initial sequence estimators \citep{dai:jone:2017, geye:1992,
  koso:2000}. Of these, the most popular are the batch means
estimators since they are easy to implement and are computationally
efficient.

The multivariate batch means estimator considers non-overlapping
batches of the Markov chain and constructs a sample covariance matrix
from the sample means of each batch. More formally, let $n = ab$ where
$a$ is the number of batches and $b$ is the batch size. For
$k = 0, \dots, a-1$, define
$\bar{Y}_k:= b^{-1}\sum_{t=1}^{b} h(X_{kb + t})$. The batch means
estimator of $\Sigma$ is,
\[
\hat{\Sigma}_{n,b}:= \dfrac{b}{a-1} \ds \sum_{k = 0}^{a-1} \left(\bar{Y}_k - \bar{\mu}_n\right) \left(\bar{Y}_k - \bar{\mu}_n \right)^T\,.
\]
The asymptotic behavior of batch means estimators has been well
studied, however small sample performance of batch means estimators
can be suspect in the presence of high correlation. Recently,
carefully constructed linear combinations of batch means estimators
have been proposed for improving finite sample performance of
estimators of $\Sigma$ \citep{liu:fleg:2018}. Particularly, for an
even $b$, they obtain the following flat-top batch means estimator
\[
\tilde{\Sigma}_n = 2 \hat{\Sigma}_{n,b} - \hat{\Sigma}_{n,b/2}\,,
\]
where $\hat{\Sigma}_{n,b/2}$ is the batch means estimator from a batch
size of $b/2$. 

Ensuring the batch means estimator (or its variants) is strongly and
mean-square consistent requires that the batch size $b$ and the number
of batches $a$ must be chosen so that both increase to infinity as
$n \to \infty$. The choice of batch size $b$ is critical to the
performance of the batch means estimator \citep{chak:etal:2019,
  fleg:jone:2010}. In particular, \cite{fleg:jone:2010} show that the
mean-squared-optimal batch size is $b \propto n^{1/3}$, where the
proportionality constant needs to be estimated separately, and its
size depends on the amount of serial correlation in the
chain. \cite{fleg:hugh:vats:2015} present a parametric method of
estimating this proportionality constant, yielding an easily
implementable optimal batch size.


\section{Stopping the simulation} 
\label{sec:stopping_mcmc}

The justification for using MCMC experiments to estimate features of
$F$ is asymptotic, but, in practice, the Monte Carlo sample size $n$
is finite.  Thus, the choice of $n$ is crucial to ensuring the
reliability of the MCMC experiment.  There are several approaches that
can be used to terminate the simulation.  The simplest is a fixed-time
procedure where the practitioner specifies the Monte Carlo sample size
before the experiment begins. In this case, we can estimate the Monte
Carlo error and, if it is too large, then run the simulation longer.
This leads to a sequential fixed-volume or fixed relative-volume
approach \citep{glyn:whit:1992} which terminates simulation at a
random time.  Discussion of these issues in the context of MCMC and
many more details about them can be found in \cite{fleg:gong:2015},
\citet{fleg:hara:jone:2008}, \citet{jone:hara:caff:neat:2006}, and
\citet{vats:fleg:jones:2019}.

We again focus on estimating a vector of expectations for the sake of
specificity.  A fruitful approach is to compare the volume of the
confidence region, $\text{Vol}(C_{\alpha}(n))$, to the generalized
variance of the target $F$, that is, $|\Lambda|$ where we recall
$\Lambda = \text{Var}_F(h(X_1))$ and $| \cdot |$ denotes the determinant.
In doing this, \citet{vats:fleg:jones:2019} show that simulating until
the effective sample size (ESS), defined as
\[
\text{ESS}:= n \left(\dfrac{|\Lambda|}{|\Sigma|} \right)^{1/p}\, ,
\]
is sufficiently large implies that the Monte Carlo error is
sufficiently small compared to the variability in the target
distribution.  ESS provides the number of independent and identically
distributed samples that will yield
the same generalized Monte Carlo error as the correlated sample.  As
discussed above, estimators of both $\Lambda$ and $\Sigma$ are
available, so that we can easily estimate it with
\[
\widehat{\text{ESS}}:= n \left(\dfrac{|\Lambda_n|}{|\Sigma_n|} \right)^{1/p}\,.
\]
Here $\Sigma_n$ is any of the estimators of $\Sigma$ discussed in Section~\ref{sec:estimating_sigma}. Notice that ESS depends on the function $h$. It is thus important to first
establish the quantities of interest before estimating ESS. When $p = 1$, the
ESS defined above is the same as the univariate ESS of 
\cite{kass:carlin:gelman:neal:1998}.

A natural question to ask is what ESS is sufficient for reliable estimation? \cite{vats:fleg:jones:2019} provide a principled cutoff for the ESS based on the relative quality of estimation. Suppose we are interested in making $100(1 - \alpha)\%$ confidence regions of $\mu_h$ using $\bar{\mu}_n$ such that the volume of the confidence region is an $\epsilon$th fraction of the variability of $h$ under $F$. Then simulation can stop the first time 
\begin{equation}
\label{eq:ess_bound}
	\widehat{\text{ESS}} \geq \dfrac{2^{2/p} \pi}{(p \Gamma(p/2))^{2/p}} \dfrac{\chi^2_{1-\alpha, p}}{\epsilon^2}:= M_{\alpha, \epsilon, p}\,.
\end{equation}
\cite{vats:fleg:jones:2019} showed that as $\epsilon \to 0$, stopping simulation according to \eqref{eq:ess_bound} will yield confidence regions with coverage probabilities converging to $1 - \alpha$. In practice, we do not check criterion~\eqref{eq:ess_bound} until after some minimum $n^*$ steps to avoid premature termination due to poor early estimates of $\Sigma$ and $\Lambda$. A reasonable choice of $n^*$ is $M_{\alpha, \epsilon, p}$, which can be calculated a priori since all quantities are known.

The lower bound in \eqref{eq:ess_bound} is in the spirit of sample
size calculations for a desired half-width of a confidence interval
for a simple test of means, where $\epsilon$ serves the same purpose
as a (relative) half-width. The relative tolerance level, $\epsilon$,
can be chosen according to the level of precision in the estimates
desired, typically $\epsilon \leq .05$.  Once $\widehat{\text{ESS}}$
reaches the cutoff, users can stop simulating since they can be
$100(1-\alpha)\%$ confident that estimates are within an $\epsilon$th level of tolerance relative to the target distribution.

The Gelman-Rubin-Brooks diagnostic of \cite{gelm:rubi:1992a} and
\cite{brooks:gelman:1998} is a popular tool for assessing convergence
in task (ii). The original diagnostic estimates $\Sigma$, by the
sample covariance of mean vectors from $m$ independent chains. Since
$m$ is often small, the original Gelman-Rubin-Brooks statistic has
high variability \citep{fleg:hara:jone:2008, vats:knudson:2018,
  veht:gel:2019}. However, a more robust statistic is achieved if the
estimators of $\Sigma$ described in Section~\ref{sec:estimating_sigma}
are used in the Gelman-Rubin-Brooks diagnostic
\citep{vats:knudson:2018}.
In addition, if $\hat{R}^p$ denotes the improved Gelman-Rubin-Brooks statistic, then \cite{vats:knudson:2018} showed that 
\[
\hat{R}^p \approx \sqrt{1 + \dfrac{1}{\widehat{\text{ESS}}}}\,.
\]
Thus, there is a direct relationship between ESS and the
Gelman-Rubin-Brooks statistic.  The above is specifically for a single
chain, while a multiple chain statistic is provided in
\cite{vats:knudson:2018}. 

The Gelman-Rubin-Brooks diagnostic suggests that the simulation be
terminated when $\hat{R}^p$ is below a pre-defined
cutoff. \cite{vats:knudson:2018} used the bound in
\eqref{eq:ess_bound} to obtain a cutoff for $\hat{R}^p$\,: simulation
stops the first time
\begin{equation} 
\label{eq:rhat_bound} 
\hat{R}^p \leq \sqrt{1 + \dfrac{1}{M_{\alpha, \epsilon, p}}}\,.  
\end{equation}
However, we recommend using ESS as it is more naturally interpretable
and has lower variability in readily available software.  Notice that
both $\hat{R}^p$ and $\widehat{\text{ESS}}$ are aimed at assessing
estimation and not ensuring a representative sample; that is, aimed at
addressing task (ii) and not task (i) from
Section~\ref{sec:introduction}.

\section{Extensions}
\label{sec:ext}

The practice of thinning (or subsampling) the Markov chain is common.
Thinning a Markov chain refers to using every $m$th observation in the
simulation.  This is often done to reduce autocorrelation, but the
resulting process is still a Markov chain with kernel corresponding to
$P^m$. \cite{geye:1992}, \cite{link:eaton:2012}, and
\cite{mac:ber:1994} showed that since this comes at the cost of the
number of usable samples, the variance of a thinned Monte Carlo
estimator is always larger than the variance of the original Monte
Carlo estimator.

There are, however, situations where thinning is worthwhile. \cite{geyer:1991}
and \cite{owen:2017}  argue that when post-processing on the raw MCMC data is
expensive, it may be computationally efficient to thin the samples. That is,
if the function $h$ is costly to evaluate relative to the time it takes to get
more samples, thinning is beneficial. Thinning can also be useful in
high-dimensional problems where storing the full MCMC output calls for large
memory requirements.  When the original MCMC sample is thinned, all output
analysis procedures then apply to the thinned MCMC sample.

Simulating multiple parallel chains is also common practice, and can
often be useful in parallel computing environments. However, multiple
short runs can be misleading and can retain large estimation bias
\citep[however, see][]{jacob:unbiased:2020}, thus we encourage users to run
each chain for as long as they would with a single chain.

Another issue that has received little attention is estimation of higher order
moments. These fit naturally into the discussion in
Section~\ref{sec:estimation_process}, but bring new practical challenges
since, for the same Monte Carlo sample size, estimation quality reduces
drastically as moments increase.

Our discussion has focused on the time-homogeneous Markov chains
encountered in standard MCMC applications. The theoretical and
practical tools required for other simulation techniques can be quite
different. For adaptive MCMC, \cite{atch:2011} provides estimators for
the asymptotic variance of the Monte Carlo
estimator. \cite{heck:over:2019} provide uncertainty quantification
for trans-dimensional MCMC methods. However, the literature for these
processes is not as rich as traditional MCMC, and would benefit from
further work.






\section{Example workflow} 
\label{sec:example}

Using an example, we now illustrate a workflow for MCMC output
analysis integrated with useful visualizations. We use a Bayesian
reliability model to assess the reliability of liquid crystal display
(LCD) projectors. We recognize the target distribution $F$, establish
the function of interest $h$, choose an appropriate MCMC sampler,
choose starting values, visually assess the quality of the MCMC
sampler using a preliminary Monte Carlo sample, implement the stopping
rules for $h$, report point estimates, and provide appropriate
graphical tools.

Consider the LCD projector data of \cite{hama:wil:ree:2008}. To test the manufacturer's claim of expected lamp life in an LCD projector being 1500 hours, identical lamps were placed in 31 projectors for various models and their time to failure was recorded. The data is presented in Table~\ref{tbl:lcd_data}.
\begin{table}
	\caption{LCD time to failure in projection hours for 31 projectors.}
\centering
	\begin{tabular}{ccccccc}
	\hline
	387 & 182 & 244 & 600 & 627 & 332 & 418\\
	300 & 798 & 584 & 660 &	39 & 274 & 174\\
	50 & 34 & 1895 & 158 & 974 & 345 & 1755 \\
	1752 & 473 & 81 & 954 & 1407 & 230 & 464\\
	380 & 131 & 1205 & & &\\ \hline
	\end{tabular}
	\label{tbl:lcd_data}
\end{table}
For $i = 1, \dots, 31$, let $t_i$ denote the observed failure time for each lamp. \cite{hama:wil:ree:2008} assumed that the $t_i$'s are a realization from,
\[
T_i \sim \text{Weibull}(\lambda, \beta)\,,
\] 
where $\lambda > 0$ is the scale parameter and $\beta >0$ is the shape parameter. Interest is in estimating the mean time to failure (MTTF) and the reliability function at $t = 1500$. Under the Weibull likelihood, the MTTF is
\[
\text{MTTF} = \lambda^{-1/\beta} \, \Gamma\left(1 + \dfrac{1}{\beta}\right)\,,
\]
and the reliability function is
\[
R(t) = \exp\left \{-\lambda t^{\beta}  \right \}\,.
\]
\cite{hama:wil:ree:2008} further assume priors
$\lambda \sim \text{Gamma}(2.5, 2350)$  and $\beta \sim \text{Gamma}(1, 1)$, where each is parameterized by a shape and rate parameter.  The density of the posterior is
\[
f(\lambda, \beta \mid T) \propto \lambda^{32.5} \beta^{31} \left( \prod_{i=1}^{31} t_i \right)^{\beta - 1} \exp \left\{ - \lambda \sum_{i=1}^{31} t_i^{\beta}\right\} \exp\{-  \beta \} \exp \{- 2350 \lambda \}\, ,
\] 
where the normalizing constant is unknown and we use component-wise MCMC methods \citep{john:jone:neat:2013} to sample from this distribution. The component $\lambda$ will be updated by a Gibbs step and $\beta$ will be updated by a Metropolis-Hastings step. The full conditional distribution of $\lambda$ is
\[
\lambda \mid \beta, T \sim \text{Gamma}\left( 33.5, 2350 + \sum_{i=1}^{31} t_i^{\beta} \right)\,.
\]
The full conditional distribution for $\beta$ is not available in closed-form, so we implement a Metropolis-Hastings step, yielding a Metropolis-within-Gibbs sampler \citep[see][]{robe:case:2013}. The proposal distribution is a $N( \cdot, .1^2)$, which yields an approximately optimal acceptance probability as suggested by \cite{rob:gel:gilks}. We update $\lambda$ first, followed by $\beta$, thus a starting value is only needed for $\beta$. We start from the MLE of $\beta$ which is approximately 1.12.

We are interested in estimating MTTF and $R(1500)$ to contest the
manufacturers' claims. Thus, the function whose posterior mean is of
interest $h$ is
\[
h(\lambda, \beta) = \left(\begin{array}{c} \lambda^{-1/\beta} \, \Gamma\left(1 + \dfrac{1}{\beta}\right) \\  \exp\left \{-\lambda t^{\beta}  \right \} \end{array} \right)\,.
\]

\begin{figure}[htb]
	\centering
	\includegraphics[width = 5in]{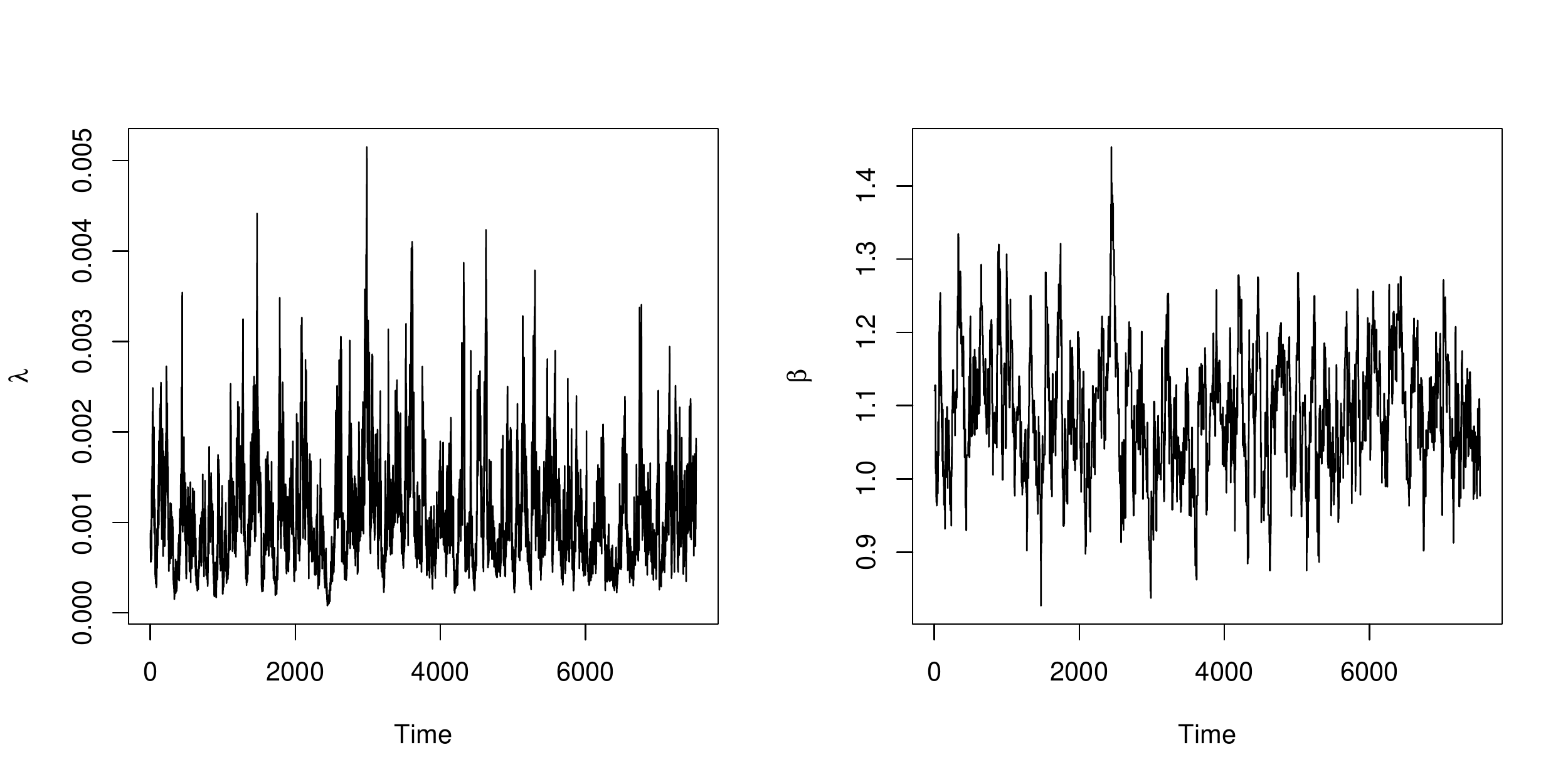}
	\caption{Trace plots of $\lambda$ and $\beta$ for an initial run of the Markov chain.}
	\label{fig:trace_lcd}
\end{figure}

Here $p = 2$ and setting the relative tolerance to be
$\epsilon = .05$ and $\alpha=.05$ yields a minimum ESS of 7529 using
\eqref{eq:ess_bound}. We first run the MCMC sampler for $n^* = 7529$  steps as
a check to see whether the sampler is exploring the state space
adequately and mixing well. Any issues with the sampler may be
addressed in such preliminary steps before a long run for a final
analysis is reported. Figure~\ref{fig:trace_lcd} shows the trace plots
of $\lambda$ and $\beta$, which indicates that there are no obvious
mixing problems.
This initial run yields $\widehat{\text{ESS}} = 1196$ which is
noticeably smaller than its cutoff. Seeing this, we run the sampler
for 92471 more steps yielding an overall sample size of 1e5. The final
$\widehat{\text{ESS}} = 11834 > 7529$ which indicates we are 95\%
confident that the simulation has terminated with a relative tolerance
smaller than $\epsilon = .05$.

\begin{figure}[H]
	\centering
	\includegraphics[width = 4.5in]{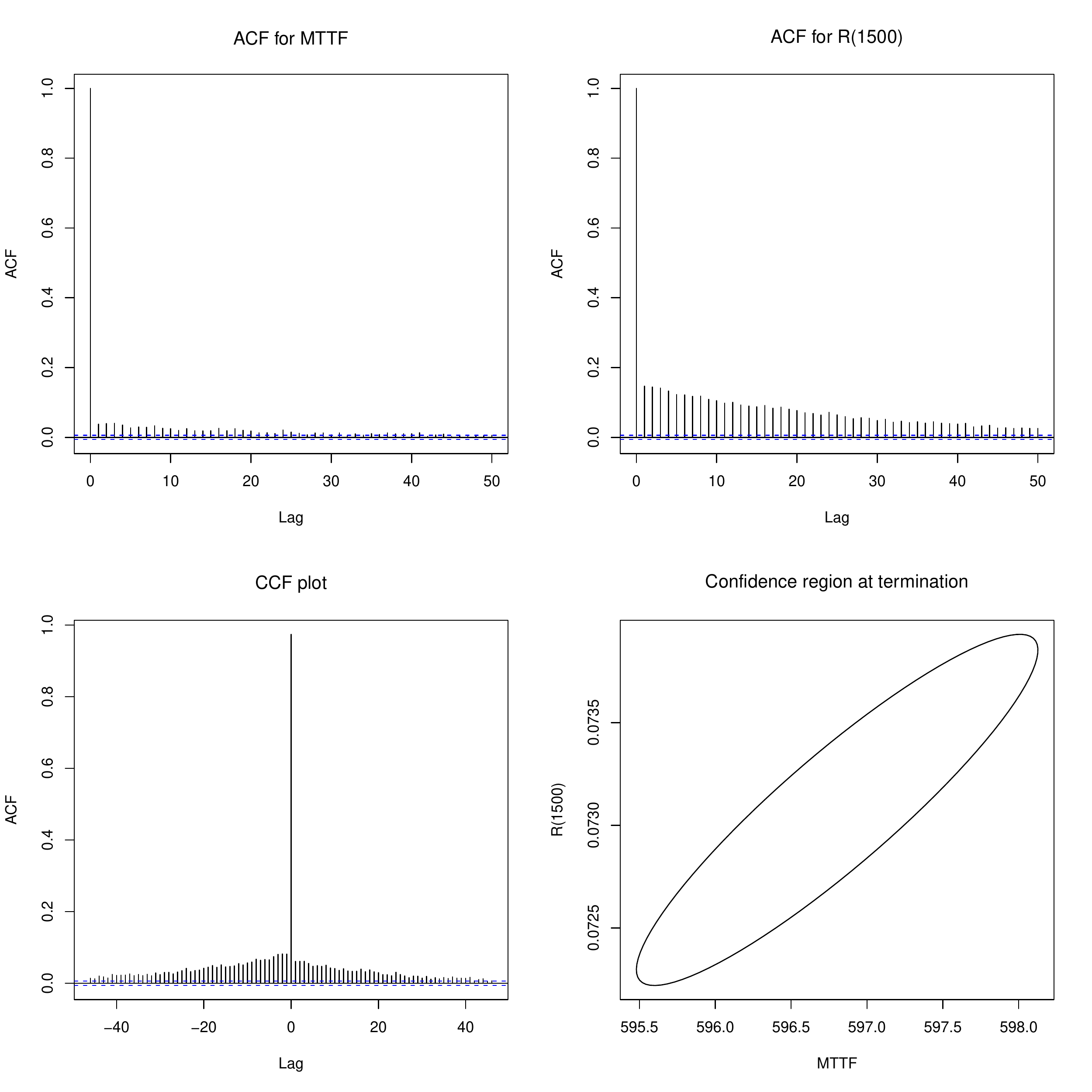}
	\caption{Autocorrelation and cross-correlation plots of MTTF
          and $R(1500)$. Also, a 95\% confidence region for the Monte
          Carlo average of MTTF and $R(1500)$. }
	\label{fig:lcd_acfs}
\end{figure}

Figure~\ref{fig:lcd_acfs} shows the autocorrelation in MTTF  and $R(1500)$ along with a cross-correlation plot and the Monte Carlo confidence region for $\bar{\mu}_n$. MTTF has little autocorrelation and $R(1500)$ has significant autocorrelation over lag 50. A multivariate analysis of the MCMC output is critically important in this example as seen by the cross-correlation plot. The cross-correlation plot at zero indicates the correlation between MTTF and $R(1500)$ in the posterior distribution. This correlation along with the lag correlations would be completely ignored by a univariate analysis. The resulting confidence region constructed using the batch means estimator and the sampling distribution in \eqref{eq:sigma_clt} captures this cross-correlation structure.

\begin{figure}[H]
	\centering
	\includegraphics[width = 6.5in]{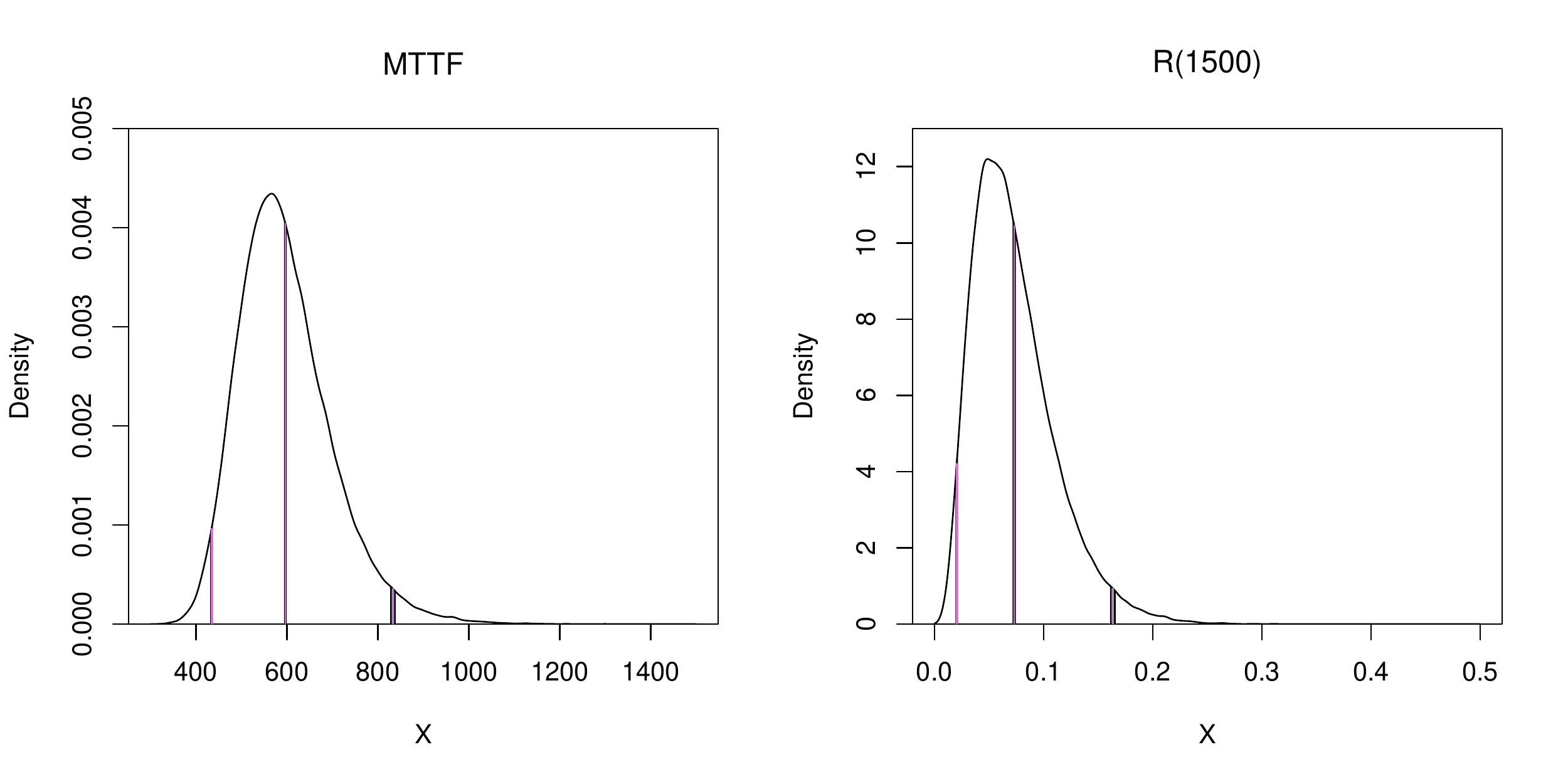}
	\caption{Marginal density plots with mean and 95\% credible
          intervals marked. Simultaneous uncertainty bands are drawn,
          but are nearly indistinguishable from the estimates.}
	\label{fig:lcd_marginal}
\end{figure}

The final estimates of MTTF and $R(1500)$ are 596.8 and 0.073
respectively. The MTTF of 596.8 is significantly far from 1500 with a
95\% credible interval of $(434, 834)$. The reliability function at a
failure time of 1500 is also low with a 95\% credible interval of
$(0.020, 0.163)$. The marginal density plots with the respective mean
and quantiles are given in Figure~\ref{fig:lcd_marginal}, along with
simultaneous uncertainty plotted for all 6 estimates using the methods
of \cite{robe:visual:2019}. The error bands around the estimates are
nearly indistinguishable from the estimates themselves, indicating a
small Monte Carlo error.

\bibliographystyle{apalike}
\bibliography{mcref}

\end{document}